\documentclass[conference, 9 pt]{IEEEtran}
\usepackage{cite}
\usepackage{amsmath,amssymb,amsfonts}
\usepackage{algorithmic}
\usepackage[ruled]{algorithm2e}
\usepackage{graphicx}
\usepackage{textcomp}
\usepackage{comment}
\usepackage{xcolor}
\usepackage{multirow}
\usepackage[a4paper, total={184mm,239mm}]{geometry}
\def\BibTeX{{\rm B\kern-.05em{\sc i\kern-.025em b}\kern-.08em
    T\kern-.1667em\lower.7ex\hbox{E}\kern-.125emX}}
\begin{document}

\title{PatchBlock: A Lightweight Defense Against Adversarial Patches for Embedded EdgeAI Devices}

\author{%
  Nandish Chattopadhyay$^{1,*}$  %
  Abdul Basit$^{1,*}$  %
  Amira Guesmi$^{1}$, %
  Muhammad Abdullah Hanif$^{1}$,\\%
  Bassem Ouni$^{2}$, %
  Muhammad Shafique$^{1}$%
  \thanks{$^{*}$These authors contributed equally to this work.}\\[0.3em]
  $^{1}$ eBRAIN Lab, Division of Engineering, New York University (NYU) Abu Dhabi, UAE\\%
  $^{2}$ DakAI, Dubai, UAE%
}


\maketitle

\begingroup
\renewcommand\thefootnote{\fnsymbol{footnote}}
\footnotetext[1]{These authors contributed equally to this work.}
\endgroup

\begin{abstract}

Adversarial attacks pose a significant challenge to the reliable deployment of machine learning models in EdgeAI applications, such as autonomous driving and surveillance, which rely on resource-constrained devices for real-time inference. Among these, patch-based adversarial attacks, where small malicious patches (e.g., stickers) are applied to objects, can deceive neural networks into making incorrect predictions with potentially severe consequences. In this paper, we present PatchBlock, a lightweight framework designed to detect and neutralize adversarial patches in images. Leveraging outlier detection and dimensionality reduction, PatchBlock identifies regions affected by adversarial noise and suppresses their impact. It operates as a pre-processing module at the sensor level, efficiently running on CPUs in parallel with GPU inference, thus preserving system throughput while avoiding additional GPU overhead. The framework follows a three-stage pipeline: splitting the input into chunks (Chunking), detecting anomalous regions via a redesigned isolation forest with targeted cuts for faster convergence (Separating), and applying dimensionality reduction on the identified outliers (Mitigating). PatchBlock is both model- and patch-agnostic, can be retrofitted to existing pipelines, and integrates seamlessly between sensor inputs and downstream models. Evaluations across multiple neural architectures, benchmark datasets, attack types, and diverse edge devices demonstrate that PatchBlock consistently improves robustness, recovering up to 77\% of model accuracy under strong patch attacks such as the Google Adversarial Patch, while maintaining high portability and minimal clean accuracy loss. Additionally, PatchBlock outperforms the state-of-the-art defenses in efficiency, in terms of computation time and energy consumption per sample, making it suitable for EdgeAI applications. 

\end{abstract}


\section{Introduction}

Adversarial attacks represent a significant challenge to the reliable deployment of deep neural network (DNN) models in real-world applications \cite{Goodfellow2015ExplainingAH}. By injecting carefully crafted perturbations into input data, these attacks exploit vulnerabilities in neural architectures and can trigger severe misclassifications. Among these, \emph{patch-based adversarial attacks} are particularly concerning due to their localized nature and practical feasibility \cite{guesmi2023physical}. Unlike distributed noise attacks, patch attacks manipulate specific regions of an image, often using conspicuous physical stickers, that can consistently deceive models with minimal adversarial effort. Such attacks have been demonstrated to undermine classification, detection, and depth estimation pipelines, raising serious risks for safety-critical domains such as autonomous driving, surveillance, and medical diagnostics \cite{guesmi2023advrain, Hu21, guesmi2024dap, googleap, lavan, guesmi2024ssap, li2021generative}.

The threat is further amplified in \emph{EdgeAI} deployments, where models operate on resource-constrained devices. These systems, integral to autonomous vehicles, IoT surveillance, and mobile robots, must balance accuracy, latency, and power consumption under tight constraints \cite{rohith2021comparative}. Conventional defense strategies, including adversarial training, model ensembles, and certified robustness \cite{adv_train, defensive_distillation, grad_reg}, impose heavy computational costs that are impractical in such environments. This motivates the need for \emph{lightweight and portable} defenses that can provide practical robustness without retraining or specialized hardware.



\begin{figure}[!htbp] 
\centerline{\includegraphics[width=1\columnwidth]{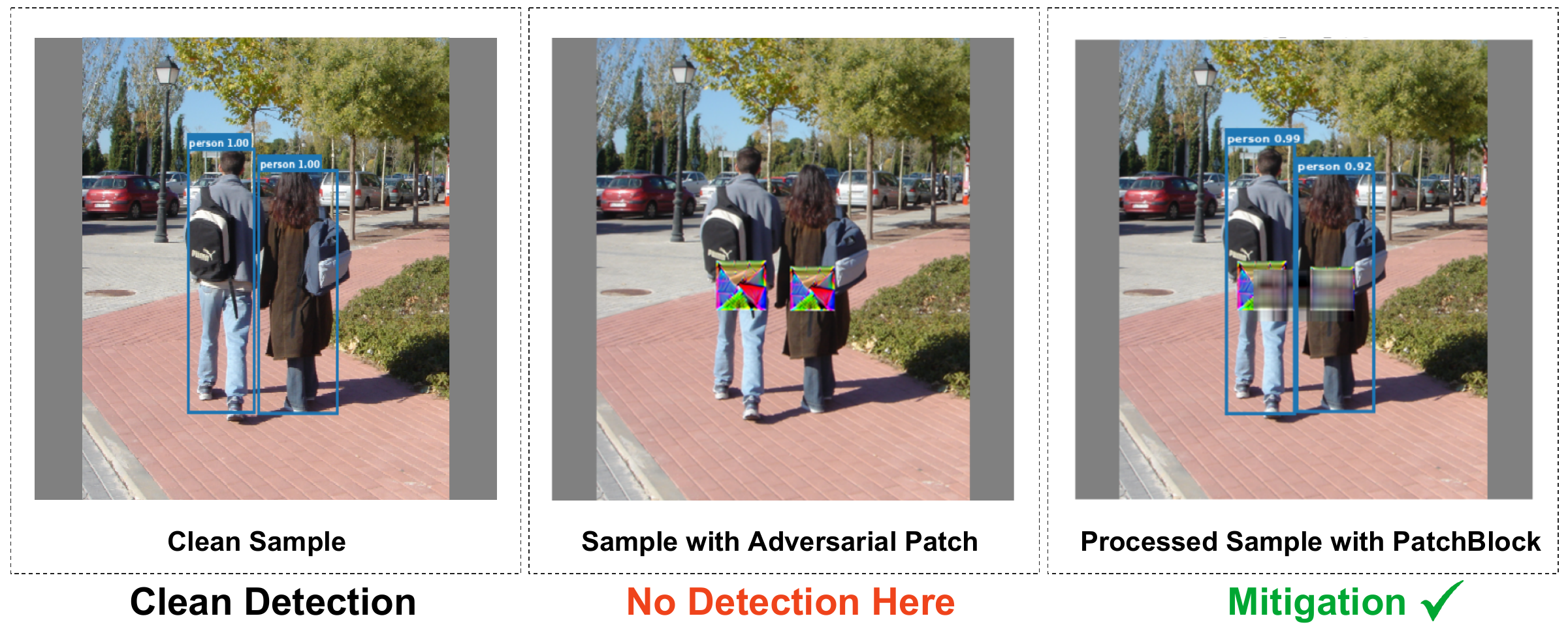}}
\caption{PatchBlock in action: detecting and mitigating adversarial patches in real time. Example shows object detection on YOLOv4 \cite{yolov4} under AdvYOLO \cite{thys2019} attack on INRIA dataset \cite{inria}.}
\label{fig:concept_1}
\end{figure} 

\begin{figure}[!htbp] 
\centerline{\includegraphics[width=1\columnwidth]{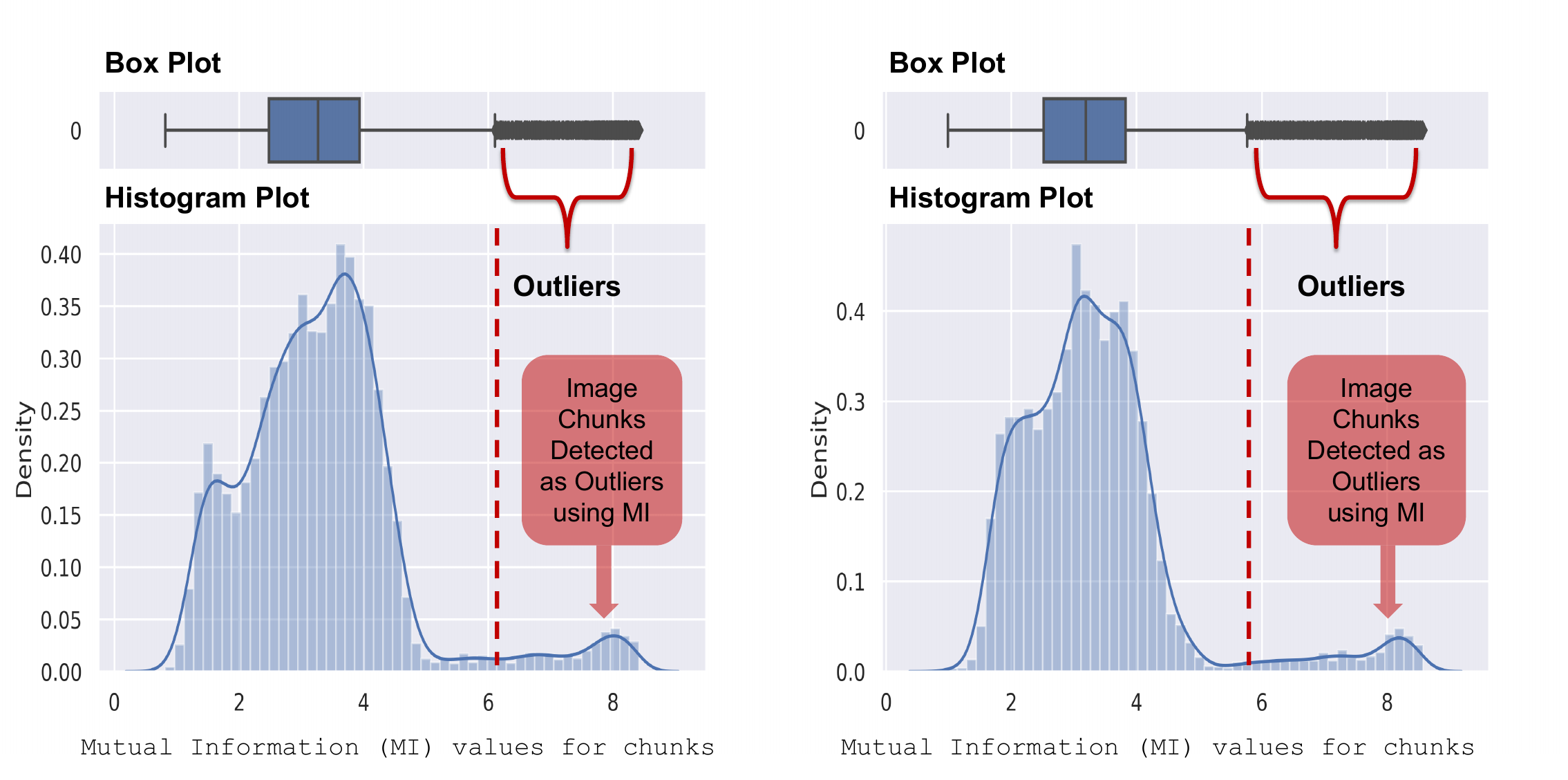}}
\caption{Key insight behind PatchBlock: adversarial patches exhibit distinct statistical distributions, captured via localized Mutual Information analysis.}
\label{fig:MI}
\end{figure}

Our study of adversarial patch behavior reveals distinctive information-theoretic signatures. Using Mutual Information (MI) \cite{MI}, we quantify localized dependencies across color channels in sliding-window chunks of an image. Regions overlapping with adversarial patches exhibit abnormally high MI values, clearly separating them from clean regions. This observation motivates our defense strategy, illustrated in Figure~\ref{fig:MI}, where adversarial regions can be efficiently identified as statistical outliers.



Building on this insight, we propose \textbf{PatchBlock}, a lightweight pre-processing framework that detects and neutralizes adversarial patches before images are fed into downstream models. PatchBlock combines:  
(i) \emph{chunking}, where inputs are partitioned into local windows;  
(ii) \emph{separating}, which applies a redesigned Isolation Forest with targeted cuts for faster convergence; and  
(iii) \emph{mitigating}, where dimensionality reduction via Singular Value Decomposition (SVD) neutralizes anomalous regions. Unlike defenses that require model retraining or GPU-bound certified verification, PatchBlock runs efficiently on CPUs and overlaps with GPU inference, preserving throughput in embedded deployments.

The novel contributions of this paper are:
\begin{itemize}
    \item We propose \textbf{PatchBlock}, a lightweight and model-agnostic adversarial defense that detects and mitigates adversarial patches by localizing affected regions and applying dimensionality reduction to neutralize their influence. PatchBlock is designed as a pre-processing block that can be retrofitted into existing pipelines without retraining.
    
    \item PatchBlock leverages statistical methods, combining localized Mutual Information analysis with Isolation Forests and Singular Value Decomposition (SVD), to identify anomalous regions efficiently. By operating primarily on CPUs and requiring no additional GPU-based computation, PatchBlock is highly suitable for deployment in resource-constrained EdgeAI environments.
    
    \item To reduce computational overhead, we redesign the Isolation Forest with a targeted-cut strategy, which accelerates convergence compared to standard implementations. In addition, we optimize Mutual Information (MI) computation method that exploits localized dependencies and parallelized processing, improving practicality for real-time defense.
    
    \item We conduct comprehensive evaluations of PatchBlock across multiple neural network models (ResNet-50 \cite{he2016deep}, VGG-19 \cite{simonyan2014very}, Vision Transformers \cite{VIT}, YOLOv4 \cite{yolov4}), datasets (ImageNet \cite{imagenet}, INRIA \cite{inria}, CASIA \cite{casia}), and adversarial patch attacks (Google Adversarial Patch (GAP) \cite{googleap}, AdvYOLO \cite{thys2019}). Results demonstrate that PatchBlock consistently improves robustness on diverse architectures and edge devices, achieving up to 77\% accuracy recovery against strong patch attacks while maintaining minimal clean performance degradation.
\end{itemize}

\begin{figure}[!htbp] 
\centerline{\includegraphics[width=1\columnwidth]{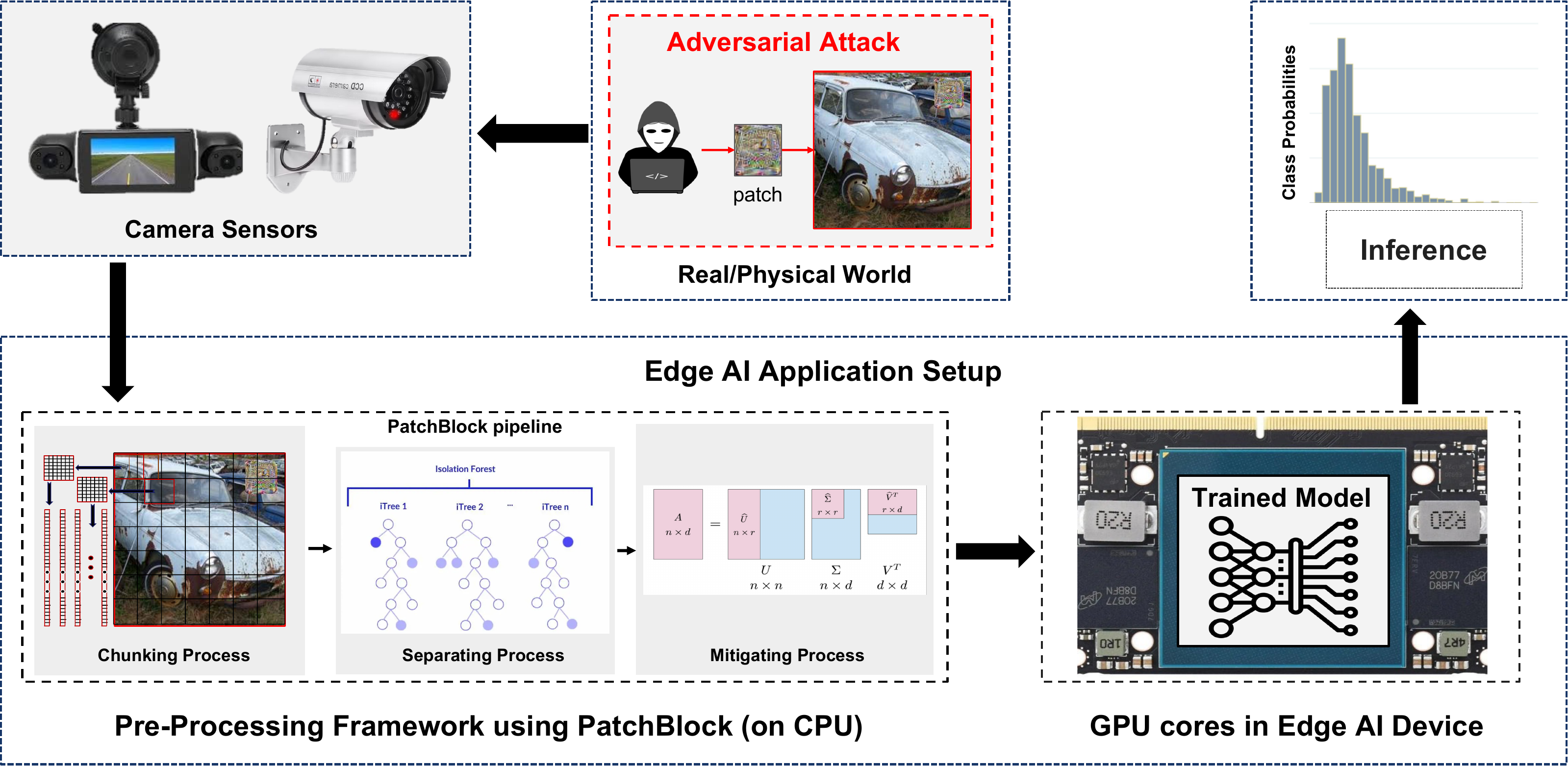}}
\caption{Overview of the PatchBlock defense pipeline, deployed between sensor inputs and downstream inference engines in embedded EdgeAI devices.}
\label{fig:concept_2}
\end{figure}

\section{Background}
In this section, we outline the theoretical underpinnings and design choices that motivate the PatchBlock defense framework.



\subsection{Threat Model: Adversarial Patch Attacks}
Adversarial patch attacks are localized perturbations where an attacker inserts conspicuous patterns (e.g., stickers) into an image to mislead deep neural networks (DNNs). These attacks have been shown to compromise classification, detection, and depth estimation pipelines, with serious implications for autonomous driving, surveillance, and medical diagnostics \cite{googleap, lavan, thys2019, guesmi2023physical, guesmi2023advrain, guesmi2024dap, guesmi2024ssap, li2021generative, PAD_CVPR2024}. We consider a white-box threat model, consistent with prior works \cite{levine2020randomized, xiang2021patchguard, naseer2019local}, where the attacker knows the model architecture and defense strategy. Although adaptive attackers may reduce robustness margins, PatchBlock demonstrates resilience against strong patch-based attacks.

\subsection{Mutual Information}
Mutual Information (MI) quantifies the amount of information shared between two random variables. For $X$ and $Y$, MI is defined as:
\begin{equation}
I(X; Y) = \sum_{x \in \mathcal{X}} \sum_{y \in \mathcal{Y}} p(x, y) \log \left(\frac{p(x, y)}{p(x)p(y)}\right)
\label{eq:MI}
\end{equation}
where $p(x,y)$ is the joint distribution and $p(x)$, $p(y)$ are the marginals. 

In images, MI captures statistical dependencies across local regions. Adversarial patches often introduce abnormal correlations, leading to MI values that deviate from those of clean regions. This provides a useful statistical signal for detection. Formally, for a patch $\mathbf{P}_{i,j}$ and its neighbors $\mathbf{N}_{i,j}$ within image $\mathbf{I}$:
\begin{equation}
I(\mathbf{P}_{i,j}; \mathbf{I}) = I\left( \mathbf{P}_{i,j}; \bigcup_{(k,l)} \mathbf{P}_{k,l} \right)
\end{equation}
Although computing global MI is computationally expensive, PatchBlock employs an efficient localized approximation that captures these anomalies while remaining deployable in real time.

\subsection{Outlier Detection}
Isolation Forest (IF) is an ensemble-based anomaly detection algorithm that recursively partitions the feature space using randomly generated decision trees \cite{IF, IFextended}. The principle is that anomalies are rare and distinct, requiring fewer partitions to isolate. In PatchBlock, adversarial patches represent such anomalies: they occupy a small region of the image while differing statistically from clean content. IF is therefore a natural choice for patch localization, as it is unsupervised, does not rely on labeled data, and remains computationally lightweight. To further improve efficiency, we introduce a \emph{targeted-cut} modification to IF. Instead of random splits, targeted cuts bias the partitioning process toward statistically informative regions, reducing tree depth and accelerating convergence. This design makes IF more practical for resource-constrained EdgeAI settings.

\subsection{Dimensionality Reduction}
Dimensionality reduction methods mitigate adversarial noise by projecting data into a lower-dimensional subspace while retaining essential information. PatchBlock uses Singular Value Decomposition (SVD) \cite{cod3, freedman}, where the ratio of singular values is used as an estimate of information retention \cite{ChattopadhyayGH24}. By attenuating anomalous components, SVD effectively suppresses patch influence while preserving clean features. While simpler operations such as zeroing out or blurring could also be applied, SVD provides a principled balance between robustness and fidelity. This choice helps PatchBlock neutralize patches without heavily degrading clean performance.

\subsection{Edge Devices and Applications}
EdgeAI platforms such as NVIDIA Jetson AGX Orin, AGX Xavier, and Orin Nano \cite{nvidia} are widely used for autonomous driving, robotics, and surveillance systems. These devices must process high-throughput sensor data under tight computational, memory, and energy constraints. Representative applications include:
\begin{itemize}
    \item \textbf{Autonomous Driving:} Real-time analysis of multimodal data (camera, LiDAR, radar) for environment perception, object detection, and navigation.
    \item \textbf{Surveillance:} On-device video analytics for anomaly detection, object tracking, and facial recognition, while reducing reliance on cloud resources.
\end{itemize}

For such use cases, a defense mechanism must improve robustness without introducing prohibitive latency or energy costs. PatchBlock is explicitly designed to meet this requirement: it executes efficiently on CPUs, overlaps with GPU inference, and introduces negligible system-level overhead.

\section{PatchBlock Algorithm}
\begin{figure*}[!htp]
\centerline{\includegraphics[width=2\columnwidth]{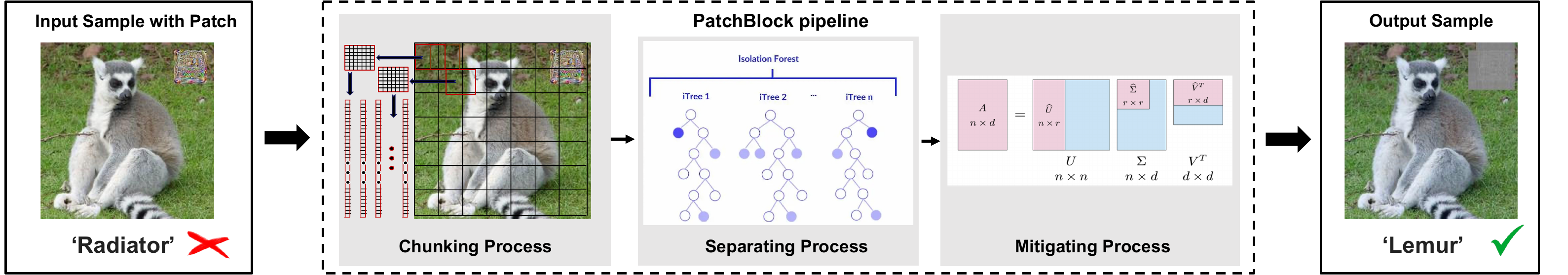}} 
\caption{PatchBlock Pipeline: The three processes of Chunking (using moving window to obtain kernels, converted to vectors), Separating (using Mutual Information and Fast Isolation Forests with targeted cuts by gradient splitting) and Mitigating (using Singular Value Decomposition). }
\label{fig:pipeline}
\end{figure*}

The implementation of the PatchBlock algorithm is carried out in three phases. The first process involves splitting up of the input image into chunks using a moving window, called the Chunking Process. This step is characterised by the size of the kernel and the stride length of the moving window. The second part deals with using Isolation Forest to detect anomalous chunks within the image that potentially contain the adversarial patch, explained in details in Algorithms \ref{forest}, \ref{tree} and \ref{gradsplit}. The final step of the process is to use dimensionality reduction with Singular Value Decomposition to mitigate the regions which may have contained the patch. This three step pipeline is outlined in Algorithm \ref{alg1}, and the corresponding schematic diagram is presented in Figure \ref{fig:pipeline}. 

\begin{algorithm}
\SetAlgoLined
\scriptsize
\caption{PatchBlock Algorithm}
\label{alg1}
$\boldsymbol{IN}:$ $I$: Unprocessed image sample, $T$: trees ensemble size in Isolation Forest, $c$: outlier score, $info$: SVD parameter,  $k$: size of kernel, $str$: stride size \\
$\boldsymbol{OUT}:$ $\hat{I}$: Processed image sample \\
/*$\boldsymbol{ChunkingProcess}$*/\\
Generate $n$ image chunks $ X \gets (x_1, \ldots, x_n)$ from image $I$, using size of kernel = $k$ and stride size = $str$ \\
/*$\boldsymbol{SeparatingProcess}$*/\\
$\textbf{Set}$ $s=0.3$ $\times$ $size(X)$ \\
$iForest = FastIsolationForest(X,T,s)$  
\\
$\boldsymbol{initialize}$ $ Y = (y_1, \ldots, y_n)$ \\
$\textbf{for}$ $i=1$ to $n$ do: \\
$y_{i} = AnomalyScore(x_{i})$, where $x_{i} \in X$  [From Algorithm \ref{forest}]\\
$\textbf{end}$ $\textbf{for}$ \\
Get $Y$ sorted in descending order \\
Pick anomalies $(z_1, \ldots, z_{r})\to Z$ as $(1-c)*n$ top $y_{i}$s    \\
/*$\boldsymbol{MitigatingProcess}$*/\\
$\textbf{for}$ $j=1$ to $r$ do: \\
$M \gets \{m_{1}, \ldots, m_{r}\}$\\
such that $m_{j} \gets x_{i}$ for matching pairs of $(i,j)$ \\
Perform Dimension Reduction using SVD as: \\
$\textbf{for}$ $k=1$ to $r$ do: \\
$\hat{m_{k}}$ $\gets$ $SVD(m_{k}, info)$ \\
Superimpose $\hat{m_{k}}$-s in place of $m_{k}$-s in image $I$ to get $\hat{I}$ \\
$\boldsymbol{return}$ $\hat{I}$
\end{algorithm}

For the purpose of detecting the anomalies, PatchBlock uses an ensemble of Fast Isolation Trees, which is a Fast Isolation Forest and is described in detail in Algorithm \ref{forest}. Specifically, we have used a significantly efficient version of the Isolation Tree, which we call Fast Isolation Tree, that uses targeted cuts on the attributes of the dataset, instead of random cuts, for quicker convergence and therefore faster detection of potential anomalies \cite{liu2018optimized}. The Fast Isolation Tree algorithm is described in Algorithm \ref{tree}, wherein it uses a function called $GradientSplit$, which is used to make the targeted cuts. 

\begin{algorithm}
\SetAlgoLined
\scriptsize
\caption{$FastIsolationForest(X,T,size)$}
\label{forest}
$\boldsymbol{IN}$: $X = (x_1, \ldots, x_n)$: input data, $T$: number of trees in the forest, $s$: sample size\\
$\boldsymbol{OUT}$: an $FastIsolationForest$ ($FastIsolationTree$s ensemble)\\
$\textbf{initialize}: FastIsolationForest$ \\
$\textbf{set}$ $height_{max} = ceiling(log_{2}$ $ s)$ \\
$\textbf{for}$ $i=1$ to $T$ do: \\
$X' \gets Sample(X,s)$ \\
$FastIsolationForest \gets FastIsolationForest \cup FastIsolationTree(X',0,height_{max})$\\
$\textbf{end}$ $\textbf{for}$ \\
$\textbf{return}$ $FastIsolationForest$
\end{algorithm}

\begin{algorithm}
\scriptsize
\caption{$FastIsolationTree(X,height,height_{max})$}
\label{tree}
$\boldsymbol{IN}$: $X = (x_1, \ldots, x_n)$: input data, $height$: height of tree, $height_{max}$: maximum height of tree, $Q$: attributes of $X$ \\
$\boldsymbol{OUT}$: a $FastIsolationTree$ \\
$\textbf{if}$ $height \geq height_{max}$ or $|X| \leq 1$, then: \\
$\textbf{return}$ $externalNode\{Size \gets |X|\}$ \\
$\textbf{else}$ \\
Randomly select $k (k \leq q$) distinct attributes from $Q$ \\
Pick out attribute $\hat{q}$ which has highest value of separability index using Algorithm \ref{gradsplit} \\
$X_{left}$ $\gets$ $\{x | x < bestX, x \in X\}$ \\
$X_{right}$ $\gets$ $\{x | x \geq bestX, x \in X\}$ \\
$\boldsymbol{return}$ $internalNode$ \\
\{ \\
$LeftTree$ $\gets$ $FastIsolationTree(X_{left}, height+1, height_{max})$ \\
$RightTree$ $\gets$ $FastIsolationTree(X_{right}, height+1, height_{max})$ \\
$attribute$ $\gets$ $bestX$ \\
$value$ $\gets$ $highestSep$ \\
\} \\
$\boldsymbol{end}$ $\boldsymbol{if}$ \\
\end{algorithm}

\subsection{Targeted Cut in Isolation Trees}
The distributions of attribute values can be used as a foundation for examining the differences between two types of instances. The underlying concept is that the cumulative discrepancy in value distributions across various attributes can indicate the extent of the difference between anomalous and normal instances. The separability of an attribute is influenced by two factors: the distance between the peaks and the spread of values within each category of instances. In this case, the values for each type of instance are averaged (denoted by $E()$) to obtain a mean value, representing the center of the values. The variance function ($Var()$) is once again used to measure the degree of dispersion in the values.  
These two metrics are used to generate the  function to calculate the separability index $Separation(X^{q},v)$, which is given as: 

\begin{equation*}
    \frac{\sqrt{(E(x|x \in X^{q}, x<v) - E(x|x \in X^{q}, x>v))^{2}}Var(X^{q})}{Var((x|x \in X^{q}, x<v)) + Var((x|x \in X^{q}, x>v))}
\end{equation*}

An approximately optimal method is proposed for searching the splitting point for a given attribute. By calculating the gradients of separability index values between adjacent attribute values, the method can help bypass attribute values unlikely to be selected as split points, thus accelerating the search process. This gradient calculating function $G$ is given as:

\begin{equation}
\label{grad}
    G = \frac{Separation(X^{q},x_{i+1}) - Separation(X^{q}, x_{i})}{x_{i+1} - x_{i}}
\end{equation}

The function which is used for updating the $step$ parameter is defined as: 

\begin{equation}
\label{step}
  step =
    \begin{cases}
      \frac{3}{1 + e^{G * log_{10}|X^{q}|}} * \frac{|X^{q}|}{100} & \text{if $G < 0$}\\
      0.7 - \frac{1.3}{1 + e^{G * log_{10}|X^{q}|}} * \frac{|X^{q}|}{100} & \text{if $G \geq 0$}
    \end{cases}       
\end{equation}

These two functions are used in the following gradient splitting algorithm for finding out the attribute with highest separability index. 

\begin{algorithm}
\scriptsize
\SetAlgoLined
\caption{$GradientSplit(X)$}
\label{gradsplit}
$\boldsymbol{IN}$: $X = (x_1, \ldots, x_n)$: input data, $X^{q}$: sorted values of $q$-th attribute\\
$\boldsymbol{OUT}$: $bestX$: attribute with highest separability index, $highestSeparation$: largest separability index value\\
$\textbf{initialize}$: $step$ as $ceiling (|X^{q}| * 0.001)$ \\
$\textbf{Let}$ $highestSeparation = Separation(X^{q}, x_{1})$ \\
$\textbf{Let}$ $bestX = (x_{1} + x_{2})/2$ and $i=0$ \\
$\textbf{while}$ $i<|X_{q}|$ do: \\
Set $i = i + step$ \\
Set $currentSeparation = Separation(X^{q}, x_{i})$ \\
$\textbf{if}$ $currentSeparation > highestSeparation$ $\textbf{then}$: \\
Set $highestSeparation = currentSeparation$ \\
Set $bestX = (x_{i} + x_{i+1})/2$ \\ 
$\textbf{end}$ $\textbf{if}$ \\
Update $step$ using equation \ref{step}, which uses equation \ref{grad}\\
$\textbf{end}$ $\textbf{while}$ \\
$\textbf{return}$ ($bestX$, $highestSeparation$)
\end{algorithm}

\subsection{Efficient Mutual Information Computation for Image Chunks}

For the task of defending against patch-based adversarial attacks, our defense mechanism involves identifying and reconstructing image chunks that exhibit anomalous behavior. A critical step in this process is the calculation of the mutual information (MI) between image chunks and the overall image, which helps in detecting regions that may have been altered by an adversarial patch. However, computing the MI between each chunk and the entire image is computationally intensive and becomes a bottleneck for real-time applications.

To address this issue, we propose an efficient method for MI computation that significantly reduces the computational load while maintaining the effectiveness of the defense mechanism. Instead of calculating the MI between each chunk and the entire image, we compute the MI between each chunk and its neighboring chunks. This approach is based on the observation that local dependencies are more relevant for detecting anomalies introduced by adversarial patches, as these patches are often localized and disrupt the local statistical properties of the image.

\subsubsection{Localized Mutual Information Computation}

We propose to compute the MI between each image chunk $\mathbf{P}_{i,j}$ and its immediate neighbors $\mathbf{N}_{i,j}$:

\begin{equation}
I_{\text{local}}(\mathbf{P}_{i,j}; \mathbf{N}_{i,j}) = \sum_{\mathbf{p} \in \mathcal{P}} \sum_{\mathbf{n} \in \mathcal{N}} p_{\mathbf{P}\mathbf{N}}(\mathbf{p}, \mathbf{n}) \log \left( \frac{p_{\mathbf{P}\mathbf{N}}(\mathbf{p}, \mathbf{n})}{p_{\mathbf{P}}(\mathbf{p}) p_{\mathbf{N}}(\mathbf{n})} \right),
\end{equation}

where $\mathcal{P}$ and $\mathcal{N}$ are the sets of possible pixel intensity values in the chunk and its neighbors, respectively.

This localized MI computation leverages the fact that natural images exhibit strong local dependencies due to spatial continuity. Anomalies introduced by adversarial patches disrupt these local dependencies, leading to significant deviations in the MI between a chunk and its neighbors.

\subsubsection{Justification for Localised MI}

The use of localized MI is theoretically sound based on the property of Markov Random Fields (MRFs) in image modeling. In an MRF, the probability of a pixel (or chunk) intensity depends primarily on its local neighborhood. Therefore, the statistical dependencies captured by the MI between a chunk and its neighbours are sufficient for detecting anomalies.

Moreover, the mutual information satisfies the property of diminishing returns with increasing neighborhood size due to the saturation of shared information. This means that adding more distant patches to the computation contributes marginally to the MI value. Mathematically, for larger neighborhoods $\mathbf{N}'_{i,j} \supseteq \mathbf{N}_{i,j}$:

\begin{equation}
I(\mathbf{P}_{i,j}; \mathbf{N}'_{i,j}) \approx I(\mathbf{P}_{i,j}; \mathbf{N}_{i,j})
\end{equation}

for sufficiently large images where the mutual information between distant chunks is minimal.

\subsubsection{Computational Efficiency}


By restricting MI computation to neighboring chunks, we reduce the complexity from $O(N^2)$ to $O(N)$, where $N$ is the number of chunks in the image. We also employ parallel processing to compute MI scores for all chunks simultaneously. Let $K$ be the number of neighbors for each chunk. The total number of MI computations is $O(N \cdot K)$, which is far less than the original $O(N^2)$ computations required when comparing each chunk to the entire image.

\section{Experimental Results}
We evaluate the effectiveness of the proposed PatchBlock defense across diverse models, datasets, and adversarial patch attacks. The goal is to demonstrate robustness, adaptability, and portability in both image classification and object detection tasks on embedded EdgeAI devices. All chosen benchmarks and attack strategies are widely used in prior work, ensuring fair comparison and reproducibility.

\subsection{Experimental Setup}
PatchBlock is deployed as a pre-processing module on CPU cores, while the underlying deep learning models run on GPU cores. This setup reflects realistic deployment scenarios where lightweight defenses operate alongside high-throughput inference engines.

\subsubsection*{Image Classification}
We evaluate classification robustness using the ImageNet dataset \cite{imagenet}, which provides diverse classes and challenging variations. The backbone models include ResNet-152 and ResNet-50 \cite{he2016deep}, VGG-19 \cite{simonyan2014very}, and Vision Transformers (ViT) \cite{VIT}, covering both convolutional and transformer-based architectures.

\subsubsection*{Person Detection}
For detection, we focus on the Person Detection task, a subdomain where adversarial patches are especially effective. We use the INRIA dataset \cite{inria}, which captures diverse real-world conditions, and the CASIA dataset \cite{casia}, which includes multiple patch instances per frame. YOLOv4 \cite{yolov4} serves as the detection backbone due to its widespread adoption and high efficiency in embedded settings.

\subsection{Adversarial Patch Attacks}
We evaluate PatchBlock against state-of-the-art adversarial patch attacks.  
\begin{itemize}
    \item \textbf{Classification:} Google Adversarial Patch (GAP) \cite{googleap} and LAVAN \cite{lavan}, which represent strong white-box attack strategies.
    \item \textbf{Detection:} AdvYOLO patch \cite{thys2019}, specifically designed to compromise YOLO-based detectors.
\end{itemize}
These attacks cover a range of patch designs and objectives, providing a rigorous test of the defense mechanism.

\subsection{EdgeAI Devices}
Experiments are conducted on two widely used embedded EdgeAI platforms:  
\begin{itemize}
    \item \textbf{NVIDIA Jetson Orin:} 12-core ARM Cortex-A78AE CPU and NVIDIA Ampere GPU with up to 2048 CUDA cores.  
    \item \textbf{NVIDIA Jetson Orin Nano:} 6-core ARM Cortex-A78AE CPU and NVIDIA Ampere GPU with up to 1024 CUDA cores.  
    \item \textbf{Lambda Tensorbook:} Intel Core i7-11800H (8 cores, 2.3–4.6 GHz), NVIDIA 3080 Ti laptop GPU with 16 GB VRAM.
\end{itemize}
PatchBlock runs exclusively on the CPU cores, while the classification and detection models are executed on the GPU. This separation demonstrates that PatchBlock introduces minimal overhead, enabling real-time deployment in resource-constrained environments.

\subsection{PatchBlock Experimental Results}

This section presents a detailed evaluation of the PatchBlock defense mechanism across diverse tasks, models, and devices, demonstrating its robustness, adaptability, and efficiency.

\begin{table}[!htbp]
\caption{Performance Analysis of PatchBlock with the Google Adversarial Patch for Image Classification Task (224x224)}
\label{tab1}
\resizebox{\columnwidth}{!}{%
\centering
\begin{tabular}{lccccc}
\hline
\multicolumn{1}{c|}{Lambda Tensorbook} & \multicolumn{1}{c|}{} & \multicolumn{1}{c|}{} & \multicolumn{1}{c|}{} & \multicolumn{1}{c}{}   \\ \cline{1-1}
\multicolumn{1}{c|}{Model} & \multicolumn{1}{c|}{\multirow{-2}{*}{\begin{tabular}[c]{@{}c@{}}Clean\\ Accuracy (\%)\end{tabular}}} & \multicolumn{1}{c|}{\multirow{-2}{*}{\begin{tabular}[c]{@{}c@{}}Adversarial\\ Accuracy (\%)\end{tabular}}} & \multicolumn{1}{c|}{\multirow{-2}{*}{\begin{tabular}[c]{@{}c@{}}Robust\\ Accuracy (\%)\end{tabular}}}  & \multicolumn{1}{c}{\multirow{-2}{*}{\begin{tabular}[c]{@{}c@{}}PatchBlock\\ Runtime (sec)\end{tabular}}} \\ \hline
\multicolumn{1}{l|}{VGG-19} & \multicolumn{1}{r|}{75.4} & \multicolumn{1}{r|}{0.45} & \multicolumn{1}{r|}{62.05}  & \multicolumn{1}{r}{0.0716} \\
\multicolumn{1}{l|}{ResNet-50} & \multicolumn{1}{r|}{79.45} & \multicolumn{1}{r|}{3.0} & \multicolumn{1}{r|}{68.65}  & \multicolumn{1}{r}{0.0738} \\
\multicolumn{1}{l|}{ViT-16} & \multicolumn{1}{r|}{83.85} & \multicolumn{1}{r|}{0.0} & \multicolumn{1}{r|}{{77.90}}  & \multicolumn{1}{r}{0.0729} \\ \hline
 &  &  &  &  &  \\ \hline
\multicolumn{1}{c|}{Jetson Orin} & \multicolumn{1}{c|}{} & \multicolumn{1}{c|}{} & \multicolumn{1}{c|}{} & \multicolumn{1}{c}{}   \\ \cline{1-1}
\multicolumn{1}{c|}{Model} & \multicolumn{1}{c|}{\multirow{-2}{*}{\begin{tabular}[c]{@{}c@{}}Clean\\ Accuracy (\%)\end{tabular}}} & \multicolumn{1}{c|}{\multirow{-2}{*}{\begin{tabular}[c]{@{}c@{}}Adversarial\\ Accuracy (\%)\end{tabular}}} & \multicolumn{1}{c|}{\multirow{-2}{*}{\begin{tabular}[c]{@{}c@{}}Robust\\ Accuracy (\%)\end{tabular}}}  & \multirow{-2}{*}{\begin{tabular}[c]{@{}c@{}}PatchBlock\\ Runtime (sec)\end{tabular}} \\ \hline
\multicolumn{1}{l|}{VGG-19} & \multicolumn{1}{r|}{77.25} & \multicolumn{1}{r|}{0.55} & \multicolumn{1}{r|}{64.80}  & \multicolumn{1}{r}{0.2052} \\
\multicolumn{1}{l|}{ResNet-50} & \multicolumn{1}{r|}{79.60} & \multicolumn{1}{r|}{3.0} & \multicolumn{1}{r|}{67.95}  & \multicolumn{1}{r}{0.2038} \\
\multicolumn{1}{l|}{ViT-16} & \multicolumn{1}{r|}{84.35} & \multicolumn{1}{r|}{0.05} & \multicolumn{1}{r|}{{ 78.10}}  & \multicolumn{1}{r}{0.1984} \\ \hline
 & \multicolumn{1}{l}{} & \multicolumn{1}{l}{} & \multicolumn{1}{l}{} & \multicolumn{1}{l}{} & \multicolumn{1}{l}{} \\ \hline
\multicolumn{1}{c|}{Jetson Orin Nano} & \multicolumn{1}{c|}{} & \multicolumn{1}{c|}{} & \multicolumn{1}{c|}{} & \multicolumn{1}{c}{}   \\ \cline{1-1}
\multicolumn{1}{c|}{Model} & \multicolumn{1}{c|}{\multirow{-2}{*}{\begin{tabular}[c]{@{}c@{}}Clean\\ Accuracy (\%)\end{tabular}}} & \multicolumn{1}{c|}{\multirow{-2}{*}{\begin{tabular}[c]{@{}c@{}}Adversarial\\ Accuracy (\%)\end{tabular}}} & \multicolumn{1}{c|}{\multirow{-2}{*}{\begin{tabular}[c]{@{}c@{}}Robust\\ Accuracy (\%)\end{tabular}}}  & \multirow{-2}{*}{\begin{tabular}[c]{@{}c@{}}PatchBlock\\ Runtime (sec)\end{tabular}} \\ \hline
\multicolumn{1}{l|}{VGG-19} & \multicolumn{1}{r|}{78.15} & \multicolumn{1}{r|}{0.7} & \multicolumn{1}{r|}{65.45}  & \multicolumn{1}{r}{0.3174} \\
\multicolumn{1}{l|}{ResNet-50} & \multicolumn{1}{r|}{78.55} & \multicolumn{1}{r|}{3.05} & \multicolumn{1}{r|}{66.0}  & \multicolumn{1}{r}{0.3177} \\
\multicolumn{1}{l|}{ViT-16} & \multicolumn{1}{r|}{83.95} & \multicolumn{1}{r|}{0.05} & \multicolumn{1}{r|}{{78.10}}  & \multicolumn{1}{r}{0.3160} \\ \hline
\end{tabular}
}
\end{table}


We evaluate PatchBlock on the ImageNet dataset using three neural network models, VGG-19, ResNet-50, and Vision Transformers (ViT-16), under the Google Adversarial Patch (GAP) attack. Table~\ref{tab1} reports clean accuracy, adversarial accuracy (with GAP applied), robust accuracy (after applying PatchBlock), and \textbf{CPU runtime} across three devices: the Lambda Tensorbook, Jetson Orin, and Jetson Orin Nano. PatchBlock substantially improved model robustness, achieving up to 77.90\% robust accuracy on ViT-16 with the Tensorbook, recovering from 0\% adversarial accuracy. Robust accuracy remained consistent across devices despite hardware differences; for example, VGG-19 achieved 64.80\% on Jetson Orin and 65.45\% on Jetson Orin Nano. Moreover, PatchBlock’s CPU runtime was well suited for resource-constrained environments, ranging from 0.0716 seconds on the Tensorbook to 0.3174 seconds on the Jetson Orin Nano for VGG-19.

\begin{table}[!htbp]
\caption{Performance Analysis of PatchBlock with the LAVAN Patch for Image Classification Task (224x224)}
\label{tab2}
\resizebox{\columnwidth}{!}{%
\centering
\begin{tabular}{lccccc}
\hline
\multicolumn{1}{c|}{Lambda Tensorbook} & \multicolumn{1}{c|}{} & \multicolumn{1}{c|}{} & \multicolumn{1}{c|}{} & \multicolumn{1}{c}{}   \\ \cline{1-1}
\multicolumn{1}{c|}{Model} & \multicolumn{1}{c|}{\multirow{-2}{*}{\begin{tabular}[c]{@{}c@{}}Clean\\ Accuracy (\%)\end{tabular}}} & \multicolumn{1}{c|}{\multirow{-2}{*}{\begin{tabular}[c]{@{}c@{}}Adversarial\\ Accuracy (\%)\end{tabular}}} & \multicolumn{1}{c|}{\multirow{-2}{*}{\begin{tabular}[c]{@{}c@{}}Robust\\ Accuracy (\%)\end{tabular}}}  & \multicolumn{1}{c}{\multirow{-2}{*}{\begin{tabular}[c]{@{}c@{}}PatchBlock\\ Runtime (sec)\end{tabular}}} \\ \hline
\multicolumn{1}{l|}{VGG-19} & \multicolumn{1}{r|}{75.65} & \multicolumn{1}{r|}{0} & \multicolumn{1}{r|}{63.9}  & \multicolumn{1}{r}{0.0758} \\
\multicolumn{1}{l|}{ResNet-50} & \multicolumn{1}{r|}{81.10} & \multicolumn{1}{r|}{6.0} & \multicolumn{1}{r|}{72.3}  & \multicolumn{1}{r}{0.0725} \\
\multicolumn{1}{l|}{ViT-16} & \multicolumn{1}{r|}{82.40} & \multicolumn{1}{r|}{5.8} & \multicolumn{1}{r|}{{73.1}}  & \multicolumn{1}{r}{0.0744} \\ \hline
 &  &  &  &  &  \\ \hline
\multicolumn{1}{c|}{Jetson Orin} & \multicolumn{1}{c|}{} & \multicolumn{1}{c|}{} & \multicolumn{1}{c|}{} & \multicolumn{1}{c}{}   \\ \cline{1-1}
\multicolumn{1}{c|}{Model} & \multicolumn{1}{c|}{\multirow{-2}{*}{\begin{tabular}[c]{@{}c@{}}Clean\\ Accuracy (\%)\end{tabular}}} & \multicolumn{1}{c|}{\multirow{-2}{*}{\begin{tabular}[c]{@{}c@{}}Adversarial\\ Accuracy (\%)\end{tabular}}} & \multicolumn{1}{c|}{\multirow{-2}{*}{\begin{tabular}[c]{@{}c@{}}Robust\\ Accuracy (\%)\end{tabular}}}  & \multirow{-2}{*}{\begin{tabular}[c]{@{}c@{}}PatchBlock\\ Runtime (sec)\end{tabular}} \\ \hline
\multicolumn{1}{l|}{VGG-19} & \multicolumn{1}{r|}{76.60} & \multicolumn{1}{r|}{0} & \multicolumn{1}{r|}{65.3}  & \multicolumn{1}{r}{0.2068} \\
\multicolumn{1}{l|}{ResNet-50} & \multicolumn{1}{r|}{79.85} & \multicolumn{1}{r|}{8.6} & \multicolumn{1}{r|}{72.05}  & \multicolumn{1}{r}{0.2007} \\
\multicolumn{1}{l|}{ViT-16} & \multicolumn{1}{r|}{84.70} & \multicolumn{1}{r|}{6.4} & \multicolumn{1}{r|}{{72.85}}  & \multicolumn{1}{r}{0.2037} \\ \hline
 & \multicolumn{1}{l}{} & \multicolumn{1}{l}{} & \multicolumn{1}{l}{} & \multicolumn{1}{l}{} & \multicolumn{1}{l}{} \\ \hline
\multicolumn{1}{c|}{Jetson Orin Nano} & \multicolumn{1}{c|}{} & \multicolumn{1}{c|}{} & \multicolumn{1}{c|}{} & \multicolumn{1}{c}{}   \\ \cline{1-1}
\multicolumn{1}{c|}{Model} & \multicolumn{1}{c|}{\multirow{-2}{*}{\begin{tabular}[c]{@{}c@{}}Clean\\ Accuracy (\%)\end{tabular}}} & \multicolumn{1}{c|}{\multirow{-2}{*}{\begin{tabular}[c]{@{}c@{}}Adversarial\\ Accuracy (\%)\end{tabular}}} & \multicolumn{1}{c|}{\multirow{-2}{*}{\begin{tabular}[c]{@{}c@{}}Robust\\ Accuracy (\%)\end{tabular}}}  & \multirow{-2}{*}{\begin{tabular}[c]{@{}c@{}}PatchBlock\\ Runtime (sec)\end{tabular}} \\ \hline
\multicolumn{1}{l|}{VGG-19} & \multicolumn{1}{r|}{75.95} & \multicolumn{1}{r|}{0} & \multicolumn{1}{r|}{63.65}  & \multicolumn{1}{r}{0.3229} \\
\multicolumn{1}{l|}{ResNet-50} & \multicolumn{1}{r|}{79.60} & \multicolumn{1}{r|}{3.25} & \multicolumn{1}{r|}{70.65}  & \multicolumn{1}{r}{0.3115} \\
\multicolumn{1}{l|}{ViT-16} & \multicolumn{1}{r|}{84.60} & \multicolumn{1}{r|}{6.5} & \multicolumn{1}{r|}{{83.20}}  & \multicolumn{1}{r}{0.3161} \\ \hline
\end{tabular}
}
\end{table}

In Table \ref{tab2}, we report the performance of PatchBlock with the LAVAN Patch for Image Classification Task on the ImageNet dataset. The trends are consistent with those of the Google Adversarial Patch.

\begin{table}[!htbp]
\caption{Performance Analysis of PatchBlock with the Adv-YOLO Patch for Object (Person) Detection Task (416x416)}
\label{tab3}
\resizebox{\columnwidth}{!}{%
\centering
\begin{tabular}{lccccc}
\hline
\multicolumn{1}{l|}{Lambda Tensorbook} & \multicolumn{1}{c|}{\multirow{2}{*}{\begin{tabular}[c]{@{}c@{}}Clean\\ Accuracy (\%)\end{tabular}}} & \multicolumn{1}{c|}{\multirow{2}{*}{\begin{tabular}[c]{@{}c@{}}Adversarial\\ Accuracy (\%)\end{tabular}}} & \multicolumn{1}{c|}{\multirow{2}{*}{\begin{tabular}[c]{@{}c@{}}Robust\\ Avg Precision (\%)\end{tabular}}}  & \multirow{2}{*}{\begin{tabular}[c]{@{}c@{}}PatchBlock\\ Runtime (sec)\end{tabular}} \\ \cline{1-1}
\multicolumn{1}{l|}{Dataset} & \multicolumn{1}{c|}{} & \multicolumn{1}{c|}{} & \multicolumn{1}{c|}{} & \multicolumn{1}{c}{} &  \\ \hline
\multicolumn{1}{l|}{INRIA} & \multicolumn{1}{r|}{100} & \multicolumn{1}{r|}{14.28} & \multicolumn{1}{r|}{92.86}  & \multicolumn{1}{r}{0.326} \\
\multicolumn{1}{l|}{CASIA} & \multicolumn{1}{r|}{100} & \multicolumn{1}{r|}{17.07} & \multicolumn{1}{r|}{100}  & \multicolumn{1}{r}{0.315} \\ \hline
 & \multicolumn{1}{l}{} & \multicolumn{1}{l}{} & \multicolumn{1}{l}{} & \multicolumn{1}{l}{} & \multicolumn{1}{l}{} \\ \hline
\multicolumn{1}{l|}{Jetson Orin} & \multicolumn{1}{c|}{\multirow{2}{*}{\begin{tabular}[c]{@{}c@{}}Clean\\ Accuracy (\%)\end{tabular}}} & \multicolumn{1}{c|}{\multirow{2}{*}{\begin{tabular}[c]{@{}c@{}}Adversarial\\ Accuracy (\%)\end{tabular}}} & \multicolumn{1}{c|}{\multirow{2}{*}{\begin{tabular}[c]{@{}c@{}}Robust\\ Avg Precision (\%)\end{tabular}}}  & \multirow{2}{*}{\begin{tabular}[c]{@{}c@{}}PatchBlock\\ Runtime (sec)\end{tabular}} \\ \cline{1-1}
\multicolumn{1}{l|}{Dataset} & \multicolumn{1}{c|}{} & \multicolumn{1}{c|}{} & \multicolumn{1}{c|}{} & \multicolumn{1}{c}{} &  \\ \hline
\multicolumn{1}{l|}{INRIA} & \multicolumn{1}{r|}{100} & \multicolumn{1}{r|}{14.28} & \multicolumn{1}{r|}{92.86}  & \multicolumn{1}{r}{0.782} \\
\multicolumn{1}{l|}{CASIA} & \multicolumn{1}{r|}{100} & \multicolumn{1}{r|}{17.07} & \multicolumn{1}{r|}{100}  & \multicolumn{1}{r}{0.764} \\ \hline
& \multicolumn{1}{l}{} & \multicolumn{1}{l}{} & \multicolumn{1}{l}{} & \multicolumn{1}{l}{} & \multicolumn{1}{l}{} \\ \hline
\multicolumn{1}{l|}{Jetson Orin Nano} & \multicolumn{1}{c|}{\multirow{2}{*}{\begin{tabular}[c]{@{}c@{}}Clean\\ Accuracy (\%)\end{tabular}}} & \multicolumn{1}{c|}{\multirow{2}{*}{\begin{tabular}[c]{@{}c@{}}Adversarial\\ Accuracy (\%)\end{tabular}}} & \multicolumn{1}{c|}{\multirow{2}{*}{\begin{tabular}[c]{@{}c@{}}Robust\\ Avg Precision (\%)\end{tabular}}}  & \multirow{2}{*}{\begin{tabular}[c]{@{}c@{}}PatchBlock\\ Runtime (sec)\end{tabular}} \\ \cline{1-1}
\multicolumn{1}{l|}{Dataset} & \multicolumn{1}{c|}{} & \multicolumn{1}{c|}{} & \multicolumn{1}{c|}{} & \multicolumn{1}{c}{} &  \\ \hline
\multicolumn{1}{l|}{INRIA} & \multicolumn{1}{r|}{100} & \multicolumn{1}{r|}{14.28} & \multicolumn{1}{r|}{92.86}  & \multicolumn{1}{r}{1.384} \\
\multicolumn{1}{l|}{CASIA} & \multicolumn{1}{r|}{100} & \multicolumn{1}{r|}{17.07} & \multicolumn{1}{r|}{100}  & \multicolumn{1}{r}{1.352} \\ \hline
\end{tabular}
}
\end{table}

PatchBlock was further evaluated for person detection on the INRIA and CASIA datasets under Adv-YOLO patch attack using the YOLO-v4 model. Table \ref{tab3} highlights clean accuracy, adversarial accuracy, robust average precision (AP), and runtime.
PatchBlock restored average precision to 92.86\% on INRIA and 100\% on CASIA, recovering from adversarial AP values as low as 14.28\%. Runtime remained within feasible limits for real-time applications, ranging from 0.326 seconds on Tensorbook to 1.384 seconds on Jetson Orin Nano for INRIA.


\begin{figure}[!htbp] 
\centerline{\includegraphics[width=0.9\columnwidth]{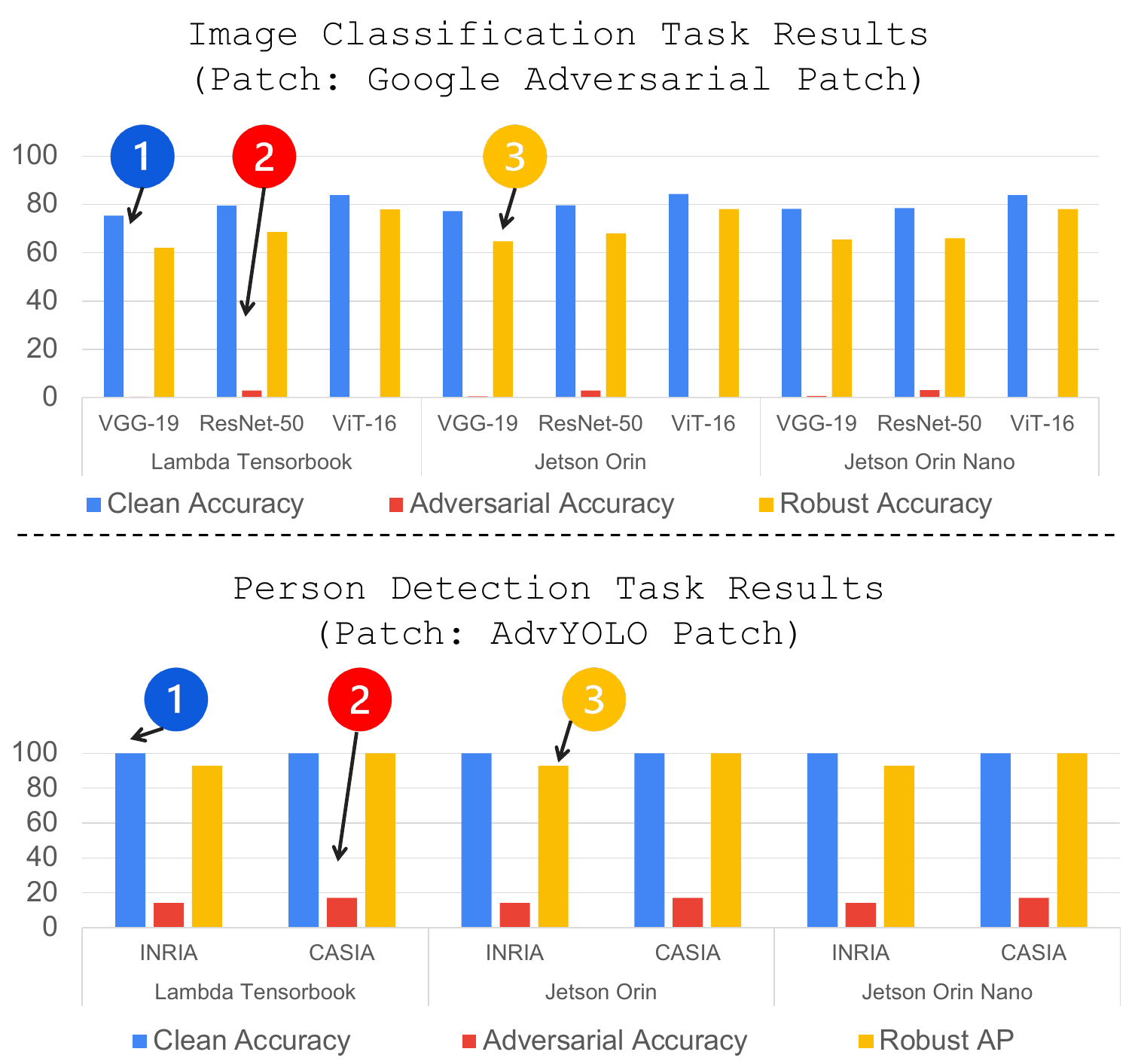}} 
\caption{PatchBlock performance across Image Classification and Person Detection tasks under adversarial patch attacks. 
Results show clean accuracy, adversarial accuracy (with GAP or AdvYOLO patches), and robust accuracy obtained {after applying PatchBlock as a pre-processing defense}. PatchBlock consistently restores accuracy across models, datasets, and devices, demonstrating its portability and effectiveness.}
\label{fig:results}
\end{figure}

\subsection{Key Findings}
We present the key findings of all the experiments in Figure \ref{fig:results}. Consistent across machine learning tasks, neural network models, adversarial patches and EdgeAI devices, we see that PatchBlock is successful in providing robustness, generating high Robust Accuracy (pointer $3$ in yellow), which is significantly higher than the Adversarial Accuracy (pointer $2$ in red), almost closing the gap on the Clean Baseline Accuracy (pointer $1$ in blue). 

\subsection{PatchBlock vs. State of the art}




To compare our work with the state-of-the-art, we review adversarial defense techniques for adversarial patches. which can be broadly categorized into two types: certified defenses and empirical defenses. Certified defenses include \textit{De-randomized Smoothing (DS)} \cite{levine2020randomized} and \textit{PatchGuard} \cite{xiang2021patchguard}. Empirical defenses include methods like \textit{Dimensionality Reduction} \cite{ChattopadhyayGH24}, \textit{Jujutsu} \cite{Jujutsu}, \textit{ODDR} \cite{chattopadhyay2023oddr}, Anomaly Detection \cite{chattopadhyay2024anomaly}, \textit{FNC} \cite{FNC} and  \textit{Localized Gradient Smoothing (LGS)} \cite{naseer2019local}.


\begin{table}[!htbp]
\caption{Performance of our proposed defense compared to four state-of-the-art defenses against GAP \cite{googleap} attack with the ResNet-50 model on the ImageNet dataset.}
\label{tab4}
\resizebox{\columnwidth}{!}{%
\centering
\begin{tabular}{l|r|l|r}
\hline
\multicolumn{1}{c|}{\multirow{2}{*}{\begin{tabular}[c]{@{}c@{}}Defense\\ Technique\end{tabular}}} & \multicolumn{1}{c|}{\multirow{2}{*}{\begin{tabular}[c]{@{}c@{}}Robust\\ Accuracy (\%)\end{tabular}}} & \multicolumn{1}{c|}{\multirow{2}{*}{\begin{tabular}[c]{@{}c@{}}Defense\\ Technique\end{tabular}}} & \multicolumn{1}{c}{\multirow{2}{*}{\begin{tabular}[c]{@{}c@{}}Robust\\ Accuracy (\%)\end{tabular}}} \\
\multicolumn{1}{c|}{} & \multicolumn{1}{c|}{} & \multicolumn{1}{c|}{} & \multicolumn{1}{c}{} \\ \hline
Localised Gradient Smoothing \cite{naseer2019local} & 53.86\% & Jujutsu \cite{Jujutsu} & 60\% \\
De-Randomised Smoothing \cite{levine2020randomized} & 35.02\% & FNC \cite{FNC} & 59.6\% \\
PatchGuard \cite{xiang2021patchguard} & 30.96\% & Anomaly Unveiled \cite{chattopadhyay2024anomaly}& 67.10\% \\
Dimensionality Reduction \cite{dac24} & 66.2\% & \textbf{PatchBlock (Ours)} & \textbf{68.65\%} \\ \hline
\end{tabular}
}
\end{table}


PatchBlock achieved 68.65\% robust accuracy, outperforming existing defenses such as Jujutsu (60\%) and Feature Norm Clipping (59.6\%). In addition to its accuracy gains, PatchBlock has lower computational requirements, making it more suitable for deployment on resource-constrained EdgeAI devices compared to more complex defenses like ODDR. The runtime results, summarized in Table~\ref{tab5}, further highlight PatchBlock’s efficiency advantage.

\noindent \textbf{Efficiency Comparison.} 
PatchBlock (PB) achieves the lowest runtime and energy cost among defenses. 
Execution time per sample: PB = 0.317\,s, ODDR = 0.431\,s, JEDI\cite{jedi} = 1.133\,s. 
Energy consumption per batch: PB = 15.83\,J, ODDR = 20.57\,J, JEDI = 48.87\,J. The energy values are estimated from measured device power (via onboard sensors on Tensorbook) integrated over time. 
Thus, PatchBlock outperforms ODDR and JEDI by a significant margin in both efficiency metrics.

\section{Discussions}
In this section, we explore the tuning of hyper-parameters for PatchBlock and analyze the impact of its optimizations on runtime efficiency.

\subsection{Hyper-Parameter Tuning for PatchBlock}

The following are the primary hyper-parameters used in PatchBlock:

\noindent\textit{Kernel Size $k$:} The kernel size determines the dimensions of the chunks created during the \textit{Chunking Process}. Larger kernel sizes reduce the number of chunks, speeding up processing but potentially reducing detection accuracy. Conversely, smaller kernel sizes provide finer granularity but increase computational overhead. Through experimentation, a kernel size of 50 pixels was found to balance accuracy and runtime effectively.

\noindent\textit{Information Retention $info$:} During the \textit{Mitigation Process}, the dimensionality reduction via SVD is applied to identified anomalous chunks. The $info$ parameter specifies the proportion of variability to retain in the reduce dimension. Retaining $85\%-90\%$ of the original information was optimal, ensuring sufficient mitigation of adversarial patches while preserving the integrity of clean data.

\noindent\textit{Outlier Score Threshold $c$:} The outlier score determines the threshold for identifying chunks as anomalous in the \textit{Separating Process} using the Isolation Forest algorithm. An optimal threshold of 1\% was selected, ensuring reliable detection of adversarial regions without excessive noise.

\subsection{Impact of PatchBlock Optimizations on Runtime}

Our experimental results demonstrate that the proposed method of using Localized Mutual Information reduces the MI computation time from approximately 250 ms to 60 ms per image on Lambda Tensorbook (RTX 3080Ti m GPU, Intel Core i7-12800H CPU), without compromising the effectiveness of the defense mechanism. This improvement enables real-time processing and makes the defense method more practical for deployment. Table \ref{tab5} reports the absolute runtime of the PatchBlock defense and the corresponding neural network model in use, across different devices, for the Image Classification task. The numbers mentioned here are for one sample. The same strategy of using batches was also used for the other task of Object Detection, since the proportions of runtime were very similar. 

\begin{table}[!htbp]
\caption{Analysis of Runtime of PatchBlock and Models across Devices}
\label{tab5}
\resizebox{\columnwidth}{!}{%
\centering
\begin{tabular}{lrrrrrr}
\hline
\multicolumn{1}{l|}{Time (sec)} & \multicolumn{2}{c|}{Lambda Tensorbook} & \multicolumn{2}{c|}{Jetson Orin} & \multicolumn{2}{c}{Jetson Orin Nano} \\ \hline
\multicolumn{1}{l|}{Models} & \multicolumn{1}{c|}{PatchBlock} & \multicolumn{1}{c|}{Model} & \multicolumn{1}{c|}{PatchBlock} & \multicolumn{1}{c|}{Model} & \multicolumn{1}{c|}{PatchBlock} & \multicolumn{1}{c}{Model} \\ \hline
\multicolumn{1}{l|}{VGG-19} & \multicolumn{1}{r|}{0.0716} & \multicolumn{1}{r|}{0.013} & \multicolumn{1}{r|}{0.2052} & \multicolumn{1}{r|}{0.0534} & \multicolumn{1}{r|}{0.3174} & 0.1125 \\
\multicolumn{1}{l|}{ResNet-50} & \multicolumn{1}{r|}{0.0738} & \multicolumn{1}{r|}{0.0179} & \multicolumn{1}{r|}{0.2038} & \multicolumn{1}{r|}{0.0613} & \multicolumn{1}{r|}{0.3177} & 0.0722 \\
\multicolumn{1}{l|}{VIT-16} & \multicolumn{1}{r|}{0.0729} & \multicolumn{1}{r|}{0.0227} & \multicolumn{1}{r|}{0.1984} & \multicolumn{1}{r|}{0.0947} & \multicolumn{1}{r|}{0.3160} & 0.0915 \\ \hline
\end{tabular}
}
\end{table}

It is note-worthy that PatchBlock runs on the CPU cores and the model itself runs on the GPU cores. To streamline this process and optimize resource utilization by minimizing the idle-time of CPU/GPU, we use a batch of 2-4 samples at a time to process with PatchBlock in the CPU cores, before feeding the defended sample to the model running in the GPU. This is motivated by the fact that the absolute runtime of the PatchBlock algorithm lies within 2-4 times the absolute runtime of the neural network models. This trend of Runtime for the PatchBlock defense and the corresponding neural network models is consistent across all machine learning tasks and devices, and the same strategy of using batches is equally useful. 





\section{Conclusions}

In this paper, we introduced PatchBlock, a lightweight pre-processing defense designed to counter patch-based adversarial attacks while remaining practical for deployment on embedded EdgeAI devices. PatchBlock operates through a three-phase pipeline—chunking, separating, and mitigating—that enables accurate detection and suppression of adversarial patches while preserving task-relevant features. The approach leverages a targeted-cut strategy within isolation trees, aggregated into an isolation forest, to localize suspicious regions, followed by dimensionality reduction to effectively mitigate the adversarial influence. Unlike adversarial patches, which require extensive training for each specific task, model, and dataset, PatchBlock is agnostic to such factors and demonstrates strong generalizability across tasks, architectures, and diverse patch attack strategies.



\section*{Acknowledgment}

This research was partially funded by the NYUAD Center for
Cyber Security (CCS), funded by Tamkeen under the NYUAD
Research Institute Award G1104 and the Technology Innovation Institute (TII) under the ”CASTLE: Cross-Layer Security for Machine Learning Systems IoT” project


\bibliographystyle{IEEEtran}
\bibliography{references}

@inproceedings{lavan,
  title={LaVAN: Localized and Visible Adversarial Noise},
  author={Dan Karmon}, 
  booktitle={International Conference on Machine Learning},
  year={2018}
}

@inproceedings{googleap,
title	= {Adversarial Patch},
author	= {Tom Brown},
year	= {2017},
URL	= {https://arxiv.org/pdf/1712.09665.pdf}
}

@article{levine2020randomized,
  title={(De) Randomized smoothing for certifiable defense against patch attacks},
  author={Levine, Alexander and Feizi, Soheil},
  journal={Advances in Neural Information Processing Systems},
  volume={33},
  pages={6465--6475},
  year={2020}
}

@misc{PAD_CVPR2024,
      title={PAD: Patch-Agnostic Defense against Adversarial Patch Attacks}, 
      author={Lihua Jing and Rui Wang and Wenqi Ren and Xin Dong and Cong Zou},
      year={2024},
      eprint={2404.16452},
      archivePrefix={arXiv},
      primaryClass={cs.CV},
      url={https://arxiv.org/abs/2404.16452}, 
}

@article{guesmi2024ssap,
  title={SSAP: A Shape-Sensitive Adversarial Patch for Comprehensive Disruption of Monocular Depth Estimation in Autonomous Navigation Applications},
  author={Guesmi, Amira and Hanif, Muhammad Abdullah and Alouani, Ihsen and Ouni, Bassem and Shafique, Muhammad},
  journal={arXiv preprint arXiv:2403.11515},
  year={2024}
}

@article{guesmi2023physical,
  title={Physical adversarial attacks for camera-based smart systems: Current trends, categorization, applications, research challenges, and future outlook},
  author={Guesmi, Amira and Hanif, Muhammad Abdullah and Ouni, Bassem and Shafique, Muhammad},
  journal={IEEE Access},
  year={2023},
  publisher={IEEE}
}

@inproceedings{xiang2021patchguard,
  title={$\{$PatchGuard$\}$: A provably robust defense against adversarial patches via small receptive fields and masking},
  author={Xiang, Chong and Bhagoji, Arjun Nitin and Sehwag, Vikash and Mittal, Prateek},
  booktitle={30th USENIX Security Symposium (USENIX Security 21)},
  pages={2237--2254},
  year={2021}
}

@inproceedings{naseer2019local,
  title={Local gradients smoothing: Defense against localized adversarial attacks},
  author={Naseer, Muzammal and Khan, Salman and Porikli, Fatih},
  booktitle={2019 IEEE Winter Conference on Applications of Computer Vision (WACV)},
  pages={1300--1307},
  year={2019},
  organization={IEEE}
}

@inproceedings{Jujutsu,
author = {Chen, Zitao and Dash, Pritam and Pattabiraman, Karthik},
title = {Jujutsu: A Two-Stage Defense against Adversarial Patch Attacks on Deep Neural Networks},
year = {2023},
isbn = {9798400700989},
publisher = {Association for Computing Machinery},
address = {New York, NY, USA},
doi = {10.1145/3579856.3582816},
booktitle = {Proceedings of the 2023 ACM Asia Conference on Computer and Communications Security},
pages = {689–703},
numpages = {15},
location = {Melbourne, VIC, Australia},
series = {ASIA CCS '23}
}

@INPROCEEDINGS{imagenet,
  author={Deng, Jia and Dong, Wei and Socher, Richard and Li, Li-Jia and Kai Li and Li Fei-Fei},
  booktitle={2009 IEEE Conference on Computer Vision and Pattern Recognition}, 
  title={ImageNet: A large-scale hierarchical image database}, 
  year={2009},
  volume={},
  number={},
  pages={248-255},
  doi={10.1109/CVPR.2009.5206848}}

@article{freedman,
  title={Statistics. 2007},
  author={Freedman, D and Pisani, R and Purves, R},
  journal={ISBN: 0-393970-833},
  year={1978}
}

@article{thys2019,
  author    = {Simen Thys and
               Wiebe Van Ranst and
               Toon Goedem{\'{e}}},
  title     = {Fooling automated surveillance cameras: adversarial patches to attack
               person detection},
  journal   = {CoRR},
  volume    = {abs/1904.08653},
  year      = {2019},
  url       = {http://arxiv.org/abs/1904.08653},
  eprinttype = {arXiv},
  eprint    = {1904.08653},
  bibsource = {dblp computer science bibliography, https://dblp.org}
}

@INPROCEEDINGS{Hu21,
  author={Hu, Yu-Chih-Tuan and Chen, Jun-Cheng and Kung, Bo-Han and Hua, Kai-Lung and Tan, Daniel Stanley},
  booktitle={2021 IEEE/CVF International Conference on Computer Vision (ICCV)}, 
  title={Naturalistic Physical Adversarial Patch for Object Detectors}, 
  year={2021},
  volume={},
  number={},
  pages={7828-7837},
  doi={10.1109/ICCV48922.2021.00775}}

@INPROCEEDINGS{inria,
  author={Dalal, N. and Triggs, B.},
  booktitle={2005 IEEE Computer Society Conference on Computer Vision and Pattern Recognition (CVPR'05)}, 
  title={Histograms of oriented gradients for human detection}, 
  year={2005},
  volume={1},
  number={},
  pages={886-893 vol. 1},
  doi={10.1109/CVPR.2005.177}}

@inproceedings{casia,
  title={A framework for evaluating the effect of view angle, clothing and carrying condition on gait recognition},
  author={Yu, Shiqi and Tan, Daoliang and Tan, Tieniu},
  booktitle={18th international conference on pattern recognition (ICPR'06)},
  volume={4},
  pages={441--444},
  year={2006},
  organization={IEEE}
}

@inproceedings{he2016deep,
  title={Deep residual learning for image recognition},
  author={He, Kaiming and Zhang, Xiangyu and Ren, Shaoqing and Sun, Jian},
  booktitle={Proceedings of the IEEE conference on computer vision and pattern recognition},
  pages={770--778},
  year={2016}
}

@article{simonyan2014very,
  title={Very deep convolutional networks for large-scale image recognition},
  author={Simonyan, Karen and Zisserman, Andrew},
  journal={arXiv preprint arXiv:1409.1556},
  year={2014}
}

@article{yolov4,
  title={Yolov4: Optimal speed and accuracy of object detection},
  author={Bochkovskiy, Alexey and Wang, Chien-Yao and Liao, Hong-Yuan Mark},
  journal={arXiv preprint arXiv:2004.10934},
  year={2020}
}

@inproceedings{jedi,
  title={Jedi: Entropy-based Localization and Removal of Adversarial Patches},
  author={Tarchoun, Bilel and Ben Khalifa, Anouar and Mahjoub, Mohamed Ali and Abu-Ghazaleh, Nael and Alouani, Ihsen},
  booktitle={Proceedings of the IEEE/CVF Conference on Computer Vision and Pattern Recognition},
  pages={4087--4095},
  year={2023}
}

@article{Goodfellow2015ExplainingAH,
  title={Explaining and Harnessing Adversarial Examples},
  author={I. J. Goodfellow and J. Shlens and C. Szegedy},
  journal={CoRR},
  year={2015},
  volume={abs/1412.6572}}

@inproceedings{guesmi2024dap,
  title={Dap: A dynamic adversarial patch for evading person detectors},
  author={Guesmi, Amira and Ding, Ruitian and Hanif, Muhammad Abdullah and Alouani, Ihsen and Shafique, Muhammad},
  booktitle={Proceedings of the IEEE/CVF Conference on Computer Vision and Pattern Recognition},
  pages={24595--24604},
  year={2024}
}

@article{guesmi2023advrain,
  title={Advrain: Adversarial raindrops to attack camera-based smart vision systems},
  author={Guesmi, Amira and Hanif, Muhammad Abdullah and Shafique, Muhammad},
  journal={Information},
  volume={14},
  number={12},
  pages={634},
  year={2023},
  publisher={MDPI}
}

@article{grad,
  author    = {Akshayvarun Subramanya and
               Vipin Pillai and
               Hamed Pirsiavash},
  title     = {Towards Hiding Adversarial Examples from Network Interpretation},
  journal   = {CoRR},
  volume    = {abs/1812.02843},
  year      = {2018},
  url       = {http://arxiv.org/abs/1812.02843},
  eprinttype = {arXiv},
  eprint    = {1812.02843},
  bibsource = {dblp computer science bibliography, https://dblp.org}
}

@inproceedings{IF,
  title={Isolation forest},
  author={Liu, Fei Tony and Ting, Kai Ming and Zhou, Zhi-Hua},
  booktitle={2008 eighth ieee international conference on data mining},
  pages={413--422},
  year={2008},
  organization={IEEE}
}

@article{IFextended,
  title={Extended isolation forest},
  author={Hariri, Sahand and Kind, Matias Carrasco and Brunner, Robert J},
  journal={IEEE transactions on knowledge and data engineering},
  volume={33},
  number={4},
  pages={1479--1489},
  year={2019},
  publisher={IEEE}
}

@article{li2021generative,
  title={Generative dynamic patch attack},
  author={Li, Xiang and Ji, Shihao},
  journal={arXiv preprint arXiv:2111.04266},
  year={2021}
}

@inproceedings{VIT,
  title={Tokens-to-token vit: Training vision transformers from scratch on imagenet},
  author={Yuan, Li and Chen, Yunpeng and Wang, Tao and Yu, Weihao and Shi, Yujun and Jiang, Zi-Hang and Tay, Francis EH and Feng, Jiashi and Yan, Shuicheng},
  booktitle={Proceedings of the IEEE/CVF international conference on computer vision},
  pages={558--567},
  year={2021}
}

@article{liu2018optimized,
  title={An optimized computational framework for isolation forest},
  author={Liu, Zhen and Liu, Xin and Ma, Jin and Gao, Hui},
  journal={Mathematical problems in engineering},
  volume={2018},
  number={1},
  pages={2318763},
  year={2018},
  publisher={Wiley Online Library}
}

@article{adv_train,
  title={Defense against adversarial attacks using feature scattering-based adversarial training},
  author={Zhang, Haichao and Wang, Jianyu},
  journal={Advances in neural information processing systems},
  volume={32},
  year={2019}
}

@inproceedings{defensive_distillation,
  title={Distillation as a defense to adversarial perturbations against deep neural networks},
  author={Papernot, Nicolas and McDaniel, Patrick and Wu, Xi and Jha, Somesh and Swami, Ananthram},
  booktitle={2016 IEEE symposium on security and privacy (SP)},
  pages={582--597},
  year={2016},
  organization={IEEE}
}

@article{grad_reg,
  title={Deep defense: Training dnns with improved adversarial robustness},
  author={Yan, Ziang and Guo, Yiwen and Zhang, Changshui},
  journal={Advances in Neural Information Processing Systems},
  volume={31},
  year={2018}
}

@inproceedings{rohith2021comparative,
  title={Comparative analysis of edge computing and edge devices: key technology in IoT and computer vision applications},
  author={Rohith, M and Sunil, Ajeet and others},
  booktitle={2021 International Conference on Recent Trends on Electronics, Information, Communication \& Technology (RTEICT)},
  pages={722--727},
  year={2021},
  organization={IEEE}
}

@inproceedings{ChattopadhyayGH24,
  author       = {Nandish Chattopadhyay and
                  Amira Guesmi and
                  Muhammad Abdullah Hanif and
                  Bassem Ouni and
                  Muhammad Shafique},
  title        = {Defending against Adversarial Patches using Dimensionality Reduction},
  booktitle    = {{DAC}},
  pages        = {222:1--222:6},
  publisher    = {{ACM}},
  year         = {2024}
}

@inproceedings{chattopadhyay2023oddr,
  title={Oddr: Outlier detection \& dimension reduction based defense against adversarial patches},
  author={Chattopadhyay, Nandish and Guesmi, Amira and Hanif, Muhammad Abdullah and Ouni, Bassem and Shafique, Muhammad},
  booktitle={Proceedings of the IEEE/CVF International Conference on Computer Vision},
  pages={22999--23008},
  year={2025}
}

@inproceedings{dac24,
  author       = {Nandish Chattopadhyay and
                  Amira Guesmi and
                  Muhammad Abdullah Hanif and
                  Bassem Ouni and
                  Muhammad Shafique},
  title        = {Defending against Adversarial Patches using Dimensionality Reduction},
  booktitle    = {{DAC}},
  pages        = {222:1--222:6},
  publisher    = {{ACM}},
  year         = {2024}
}

@inproceedings{FNC,
  title={Defending against universal adversarial patches by clipping feature norms},
  author={Yu, Cheng and Chen, Jiansheng and Xue, Youze and Liu, Yuyang and Wan, Weitao and Bao, Jiayu and Ma, Huimin},
  booktitle={Proceedings of the IEEE/CVF International Conference on Computer Vision},
  pages={16434--16442},
  year={2021}
}

@article{MI,
  title={Estimating mutual information},
  author={Kraskov, Alexander and St{\"o}gbauer, Harald and Grassberger, Peter},
  journal={Physical Review E—Statistical, Nonlinear, and Soft Matter Physics},
  volume={69},
  number={6},
  pages={066138},
  year={2004},
  publisher={APS}
}

@inproceedings{nvidia,
  title={Nvidia orin system-on-chip},
  author={Ditty, Michael},
  booktitle={2022 IEEE Hot Chips 34 Symposium (HCS)},
  pages={1--17},
  year={2022},
  organization={IEEE Computer Society}
}

@inproceedings{cod3,
  title={Robust Perception for Autonomous Vehicles using Dimensionality Reduction},
  author={Garg, Shivam and Chattopadhyay, Nandish and Chattopadhyay, Anupam},
  booktitle={2022 IEEE International Conference on Trust, Security and Privacy in Computing and Communications (TrustCom)},
  pages={1516--1521},
  year={2022},
  organization={IEEE}
}

@article{chattopadhyay2024anomaly,
  title={Anomaly Unveiled: Securing Image Classification against Adversarial Patch Attacks},
  author={Chattopadhyay, Nandish and Guesmi, Amira and Shafique, Muhammad},
  journal={arXiv preprint arXiv:2402.06249},
  year={2024}
}

\end{document}